\documentclass[11pt,a4 paper, double]{article} 
\usepackage{amsmath,amscd,amsfonts, graphics, color}
 \usepackage{amssymb, latexsym, framed, rotating} 
\usepackage{setspace} 
 \usepackage{authblk}
\makeatletter 
 \renewcommand\@biblabel[1]{#1} 
\makeatother %
\makeatletter
\let\@fnsymbol\@arabic
\makeatother\usepackage{comment}
\usepackage{lipsum}
\usepackage{fullpage}
\usepackage[utf8]{inputenc}
\usepackage{graphicx}
\graphicspath{ {./docu_pictures/} }
\usepackage{amssymb}
\usepackage{forest}
\usepackage{lscape}
\usepackage{fancyvrb}
\usepackage{amsmath, amsthm}
\usepackage{multirow}
\usepackage{rotating}
\usepackage{spverbatim}
\usepackage{tikz}
\usetikzlibrary{trees}
\usepackage[english]{babel}

\usepackage{geometry}
\usepackage{parskip}

\usepackage{url}

\usepackage{natbib}
\usepackage{framed}
\makeatletter
\newenvironment{kframe}{%
 \def\at@end@of@kframe{}%
 \ifinner\ifhmode%
  \def\at@end@of@kframe{\end{minipage}}%
  \begin{minipage}{\columnwidth}%
 \fi\fi%
 \def\FrameCommand##1{\hskip\@totalleftmargin \hskip-\fboxsep
 \colorbox{shadecolor}{##1}\hskip-\fboxsep
     \hskip-\linewidth \hskip-\@totalleftmargin \hskip\columnwidth}%
 \MakeFramed {\advance\hsize-\width
   \@totalleftmargin\z@ \linewidth\hsize
   \@setminipage}}%
 {\par\unskip\endMakeFramed%
 \at@end@of@kframe}
\makeatother

\definecolor{shadecolor}{rgb}{.97, .97, .97}
\definecolor{messagecolor}{rgb}{0, 0, 0}
\definecolor{warningcolor}{rgb}{1, 0, 1}
\definecolor{errorcolor}{rgb}{1, 0, 0}
\newenvironment{knitrout}{}{} 

\usepackage{alltt}

\title{EgoCor: an R package to facilitate the use of exponential semi-variograms for modelling the local spatial correlation structure in social epidemiology}

\author[1,*]{Julia Dyck}
\author[1]{Jan-Ole Koslik}
\author[1,2]{Odile Sauzet}
\affil[1]{Department of Business
Administration and Economics, Bielefeld University, Bielefeld, Germany;}
\affil[2]{Department of Epidemiology \& International Public Health, Bielefeld School of Public Health (BiSPH), Bielefeld University, Germany}
\affil[*]{Correspondence: Julia Dyck, Department of Business
Administration and Economics, Bielefeld University, Bielefeld, Germany, email: j.dyck@uni-bielefeld.de}
\date{}

\begin{document}
\maketitle

\abstract{
  As an alternative to using administrative areas for the evaluation of small-area
health inequalities, Sauzet et al.\ suggested to take an ego-centred approach and
model the spatial correlation structure of health outcomes at the individual level.
Existing tools for the analysis of spatial data in R might appear too complex to
non-specialists which could limit the use of the approach.

We present the R package EgoCor which offers a user-friendly interface
displaying in one function a range of graphics and tables of parameters to
facilitate the decision making about which exponential parameters fit best either raw
data or residuals. This function is based on the functions of the R package
gstat. Moreover, we implemented a function providing the measure of uncertainty proposed by Dyck
and Sauzet.

With the R package EgoCor the modelling of spatial correlation structure of
health outcomes or spatially structured predictors of health with a measure of uncertainty is made available to non-specialists.
}


\section{Background}\label{sec:intro}

The last 20 years have seen an increase in interest in the study of associations between neighbourhood characteristics and health. Most quantitative studies are based on neighbourhood defined as disjoint administrative spatial units and non-measured spatial effects are estimated via multilevel models. However, there are limits to this strategy which have been discussed in the literature \citep{chaix2009neighbourhoods, van2012neighbourhood}.

An alternative approach is to conceptualize neighbourhood as ego-centred such that every individual has its own neighbourhood centred on the place of residence. Such a neighbourhood must have relevance for the health outcome of interest. Some studies mentioned the use of the spatial correlation structure of health outcomes as an alternative to estimating the correlation within administrative areas \citep{chaix2005comparison}. This idea has been advanced by \citet{sauzet2021} by proposing to use the parameters of an exponential model for the semi-variogram of health outcomes to assess this correlation structure quantitatively. 

Semi-variogram models provide a measure of the presence of unmeasured spatial effects on health. They have the advantage of being a good fit for the empirical data as well as for the following concept: if the place of residence affects health, then health outcomes of neighbours are correlated and this correlation evanesces with increasing distance between neighbours. 
Another application of such an approach in social epidemiology presented by \citet{finne2024} is to use semi-variogram models to obtain ego-centred indicators of neighbourhood using kriging. These in turn can be used as predictors of health outcomes.

The procedure of investigating the spatial correlation includes first estimating an empirical semi-variogram based on the data available and then fitting an exponential parametric model to this semi-variogram. Here, only small distances are considered to model the local correlation structure, i.e. the correlation of a health outcome between immediate neighbours.

The empirical semi-variogram is not unique as it is based on the choice of the maximal distance between observations (all pairs of observations which are further apart than the maximal distance are not used to estimate the semi-variogram) and the number of bins (distance intervals for pairs of observations for which one point of the semi-variogram will be estimated).  If the number of observations is limited and only a small proportion of the variance is spatially structured, then the ability of fitting an exponential model to the semi-variogram may be strongly dependent on the mentioned meta parameters. Therefore, the fit of the estimated exponential model must be evaluated visually with a comparative evaluation of a range of maximal distances and numbers of bins and by applying a measure of uncertainty to the fitted models.

A variety of software implementations to fit semi-variogram models in R is already available, but the range of modelling possibilities can be a deterrent to its wider use by social epidemiologists with limited experience with the analysis of spatial data.
Moreover, it remains presently difficult to obtain measures of uncertainty for parametric semi-variogram models. As far as we know the only method that has been implemented in current R packages is the BRISC method by \cite{saha2018brisc} provided in the R package {BRISC} \citep{BRISCpkg} which cannot be used given geo-referenced survey data due to the large number of data points, but limited number of pairs at small distances \citep{dyck2023parameter}. \citet{dyck2023parameter} have investigated how to modify an existing bootstrap approach by \citet{olea2011generalized} to make it reliable in the context of health data surveys (sparse local data defined as a small number of pairs at small distances, and large overall sample size) and propose a filtered bootstrap estimate for the standard error of parameter estimates of an exponential semi-variogram model implementable in R.

The aim of the R package {EgoCor} \citep{EgoCor} is to offer a user-friendly interface to obtain graphics and tables of parameters for a range of fitted semi-variogram models in one function in order to facilitate the decision making about which exponential semi-variogram model parameters fit the data best. In addition, it provides a function to calculate a measure of uncertainty not available until now. Our package is based on the functions of the R package {gstat} \citep{R, Pebesma1998} and on the work of \cite{dyck2023parameter}. 

In the subsequent section, we summarize the basic concepts of semi-variogram modelling and the filtered bootstrap for semi-variogram model parameter standard error estimation (see Section \ref{sec:models}). This is followed by an introduction to all functions provided in the package accompanied by an illustrative analysis demonstrating the practical application of EgoCor (see Section \ref{sec:package}).

\section{Statistical Methods} \label{sec:models}

In this section, we recall the statistical models and algorithms implemented in the EgoCor package.
For the interested reader, more detailed sources are mentioned.

\subsection{Semi-variogram model}

We provide some basic concepts on semi-variogram modelling. For a more detailed description we refer to \cite{schabenberger2017statistical}. In the sequel, we assume that the spatial process is isotropic and second-order stationary. Under those conditions, the correlation $C(h)$ between health
outcomes $Z(s)$ and $Z(s')$ (or the residuals of a regression model) at locations $s$ and $s'$ depends only on the (lag) distance  $h =  ||s-s'||$ between those observations \citep{schabenberger2017statistical}.

The semi-variogram $\gamma(h)$ for a positive lag $h$ is defined as
$$\gamma(h)=\frac{1}{2}Var[Z(s)-Z(s')].$$

The covariance function $C(h)$ under some regularity conditions is given by
$$C(h)=c_0+\sigma^2_0-\gamma(h)$$
where $\sigma^2_0+c_0=Var[Z(s)]$ is the variance of the health outcome and the nugget effect $c_0$ is the value of the semi-variogram when the distance between two observations tends to zero. The parameter $\sigma_0^2$ is called the partial sill.

Matheron's estimator provides an unbiased estimator for the empirical semi-variogram at distance $h$ between two observations \citep{matheron1962traite}

 	 $$\hat\gamma(h)=\frac 1 {2|N(h)|}  \sum_{(s,s' )\in N(h))}\{Z(s )-Z(s')\}$$

where $N (h)$ is the set of all observations lagging at distance $h$.
This empirical semi-variogram can be estimated from the data for a finite number of lag bins.

To provide a smooth estimate, an exponential model of the form
	\begin{align*}\label{formula_exp_sv_model}
		\gamma_{exp}(h) =
		\begin{cases}
			c_0 + \sigma_0^2 \Big(1 - \exp\big(- \frac{h}{\phi}\big)\Big) \;\; & \text{for}\;\; h > 0, \\
			0 \;\; & \text{for} \;\; h = 0 .
		\end{cases}
	\end{align*}
can be fitted to the values of the empirical semi-variogram resulting in parameter estimates $\hat{c}_0, \hat{\sigma}_0^2$ and $\hat{\phi}$.

From the fitted model several statistics can be calculated: 
The practical range (distance above which the correlation between observations is less than 5\% of the total variance) for this model is given by
 $$H=\hat{\phi} \log \left ( \frac{\hat{\sigma}_0^2}{0.05(\hat{c}_0+\hat{\sigma}_0^2)}\right ).$$

The relative structured variability (RSV) calculated as partial sill divided by total variance is a measure of the degree of spatial structure:
$$RSV=\frac{\hat{\sigma_0}^2}{\hat{c_0} + \hat{\sigma_0}^2}.$$

The relative bias (RB) between the estimated variance according to the model and the sample variance of the health outcome is obtained as
$$ RB = \frac{\hat{c_0} + \hat{\sigma_0}^2}{\widehat{Var[Z(s)]}}. $$

\subsection{Filtered bootstrap standard errors}
The filtered bootstrap standard errors provide a way of uncertainty quantification for the parameters of the exponential semi-variogram model.
The algorithm used is based on
the generalized bootstrap method explained in \cite{olea2011generalized} and \cite{genboot2012} and was adapted in \cite{dyck2023parameter} to the characteristics of population data, i.e. large sample sizes overall, but low number of pairs at small distances. We briefly recall the steps of the filtered bootstrap algorithm:

\begin{enumerate}
\setcounter{enumi}{-1}
\item \textbf{Exponential semi-variogram model:} An empirical semi-variogram is fitted to the original spatial dataset to which an exponential model is fitted. This provides the parameter estimate $\mathbf{\hat{\theta}}$ for the true parameter vector $\mathbf{\theta}$ with $(\theta_1, \theta_2, \theta_3) = (c_0, \sigma^2_0, \phi)$.

	\item \textbf{Normal score transformation:} The data vector $\mathbf{z} = (z_1,\dots, z_N)^t$ is mapped into a Gaussian space by the empirical normal score transformation function $\varphi$.
Consequently, $\mathbf{y}= \varphi(\mathbf{z})$ is a realization vector of a standard normal random variable \citep{GSLIB}.
	\item \textbf{Exponential semi-variogram model for transformed data:} An empirical semi-variogram and exponential semi-variogram model $\tilde{\gamma}_{\text{exp}}$ are fitted  to the transformed data $\mathbf{y}$ combined with the raw data's geo-coding providing the parameter estimate $\mathbf{\tilde{\theta}}$.

	\item \textbf{Covariance estimation:} Making use of the exponential semi-variogram model characterized by $\mathbf{\tilde{\theta}}$, the covariance between two data points $z_i$ and $z_j$ is calculated based on the Euclidean distance $d_{ij}$ between these two points:
$$ c_{ij} = \tilde{c_0} + \tilde{\sigma_0^2} - \tilde{\gamma}_{\text{exp}}(d_{ij}). $$
The covariance matrix with entries $c_{ij}$ is denoted as $\mathbf{C}$.

	\item \textbf{Decorrelation of the data:} The decomposition of the covariance matrix $\mathbf{C}$ into the product of a lower and triangular matrix $\mathbf{L}$ and its transpose is obtained by the Cholesky decomposition algorithm as
$\mathbf{C} = \mathbf{L}\mathbf{L}^t.$

This decomposition is used to remove the correlation structure within the sample $\mathbf{y}$. The resulting vector $\mathbf{x} = \mathbf{L}^{-1}\mathbf{y}$
contains independent and identically distributed, hence uncorrelated values \citep{Solow1985}.

	\item \textbf{Classical bootstrap:} Sampling with replacement from $\mathbf{x}$ leads to a bootstrap sample $\mathbf{x^*}$ of the same size as the original spatial dataset.
	\item \textbf{Recorrelation:} The resample $\mathbf{x^*}$ reinherits the correlation structure by applying the inverse operation of step 4, that is
	$
	\mathbf{y^*} = \mathbf{Lx^*}.
	$
	\item \textbf{Normal score back transformation:} The back transformation of $\mathbf{y^*}$ to the attribute space through the inverse normal score function is obtained as
	$
	\mathbf{z^*} = \varphi^{-1}(\mathbf{y^*}).
	$

	\item \textbf{Analysis of the bootstrap sample:} An exponential semi-variogram model is estimated based on $\mathbf{z^*}$ combined with the original coordinates providing an estimate $\mathbf{\theta^*}$.

	\item \textbf{Filtering:} A check-filter-based test is applied to the bootstrap estimate $\mathbf{\theta^*}$ to indicate whether the exponential semi-variogram fitting algorithm did converge.
	If
$$
c_{0}^* + \sigma_{0}^{2*} > \tau  \widehat{Var(\mathbf{z})},
	$$
i.e. if the variance indicated by the model exceeds the estimated sample variance times the threshold factor $\tau$, the bootstrap estimate is discarded, otherwise it is saved. Within the {EgoCor} package the threshold is set to $\tau = 3$ by default as in a simulation study it was found to provide the best results with respect to the standard error estimates \citep{dyck2023parameter}.

	\item \textbf{Repetition:} The steps 5 to 9 are repeated until a set of $B$ bootstrap estimates ${\{ \theta^*_b \}}_{b = 1,\dots, B}$ has aggregated.

	\item \textbf{Parameter standard error estimation:} Based on the set of repeatedly estimated parameters ${\{ \theta^*_b \}}_{b = 1,\dots, B}$, estimates of the parameter standard errors are obtained by the empirical standard deviation
$$ \widehat{se(\theta_j)}
= sd(\theta^*_j)
= \sqrt{ \frac{1}{B-1} \sum_{b=1}^{B}
\Big\{ \theta_{b j}^* - \overline{\theta_j^*} \Big\}^2 },
$$
with $j = 1,...,3$ referring to the three parameters $c_0,\; \sigma^2_0$ and $\phi$.

\end{enumerate}




\section{The EgoCor package}\label{sec:package}

The EgoCor package is available from the Comprehensive R Archive Network (CRAN) at \url{https://cran.r-project.org/web/packages/EgoCor} \citep{EgoCor}. A development version and source code can be found on \url{https://github.com/julia-dyck/EgoCor}.

In this section, we describe the functions provided by the package EgoCor and subsequently apply them to the simulated dataset \texttt{birth} to demonstrate their practical use.

\subsection{Dataset}

The simulated dataset \texttt{birth} is provided with the package {EgoCor}. The dataset is based on the spatial distribution of real birthweight data \citep{Spalleke018398} and contains eight variables for 903 births:

\begin{itemize}
\item \texttt{x}: x-coordinate in meters for a fictive Cartesian coordinate system,
\item \texttt{y}: y-coordinate in meters for a fictive Cartesian coordinate system,
\item \texttt{birthweight}: birthweight in grams,
\item \texttt{primiparous}: first pregnancy (1) or subsequent pregnancy (0),
\item \texttt{datediff}: number of days to due date,
\item \texttt{bmi}: BMI of the mother at first medical appointment,
\item \texttt{weight}: weight of the mother at first medical appointment,
\item \texttt{inc}: income quintile (0, 1, 2, 3, 4).
\end{itemize}

\subsection{Functions}

We use the \texttt{birth} dataset to illustrate the following functions:
\begin{itemize}
\item \texttt{coords.plot()}: for graphical description of locations,
\item \texttt{distance.info()}: for descriptive information about distances between observations,
\item \texttt{vario.reg.prep()}: to model the spatial correlation structure of residuals of a regression model,
\item \texttt{vario.mod()}: to fit exponential models to semi-variograms with graphical presentation,
\item\texttt{par.uncertainty()}: to obtain bootstrap standard errors for the parameters of the exponential semi-variogram model.
\end{itemize}

\subsection{Data exploration}

The data format required by the {EgoCor} functions is either a data frame or a matrix.

The first three columns of the data frame or matrix should be ordered the following way:
\begin{itemize}
    \item 1st column: x-coordinate in meters for a Cartesian coordinate system,
    \item 2nd column: y-coordinate in meters for a Cartesian coordinate system, 
    \item 3rd column: outcome of interest.
\end{itemize}
Other columns will be ignored. A message appears following the output of the function \texttt{vario.mod()} recalling the required order for the variables.

\begin{figure}[t]
\centering
\begin{knitrout}
\begin{kframe}
\begin{alltt}
coords.plot(birth)
\end{alltt}
\end{kframe}
\includegraphics[width=\textwidth]{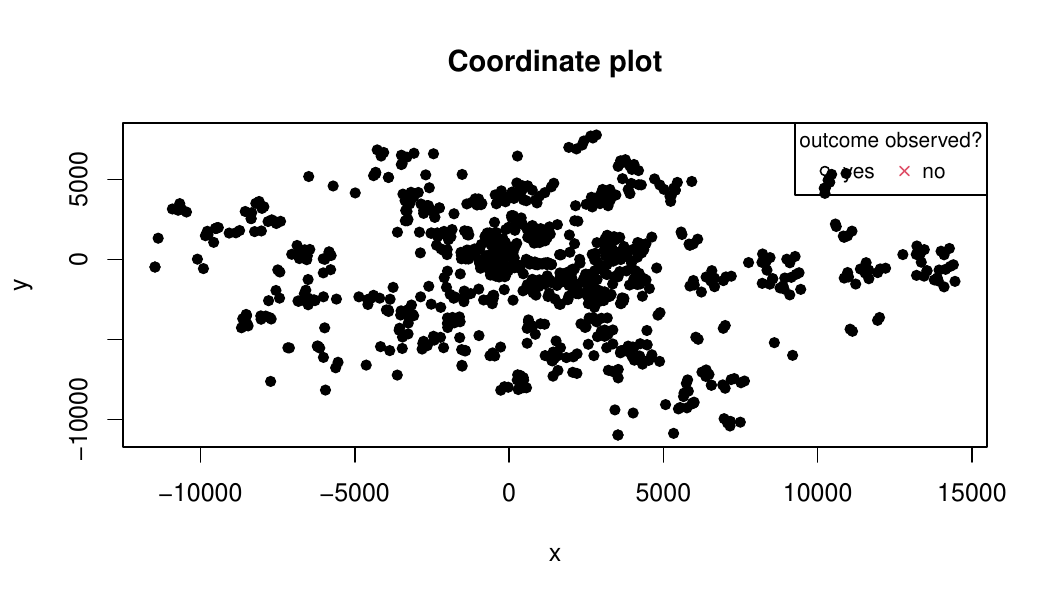} 
\end{knitrout}
\caption{\label{birth} Coordinates plot for outcome \texttt{birthweight}}
\end{figure}

The function \texttt{coords.plot()} provides a simple visualization of the locations on a two-dimensional map and indicates whether the outcome is observed (by a black circle) or missing (by a red x) at a specific location.
The purpose of this function is to investigate the spatial distribution of observations and to uncover potential spatial patterns in the distribution of missing values in the outcome of interest or covariates. Figure \ref{birth} displays the location of the observations. For the outcome \texttt{birthweight} there are no missing values.

\begin{figure}[t]
\begin{knitrout}
\begin{kframe}
\begin{alltt}
x <- sample(1:903, 30)
datainc <- birth[, c(1:2, 8)]
datainc[x, 3] <- NA
coords.plot(datainc)
\end{alltt}
\end{kframe}
\end{knitrout}
\centering
\includegraphics[width=\textwidth]{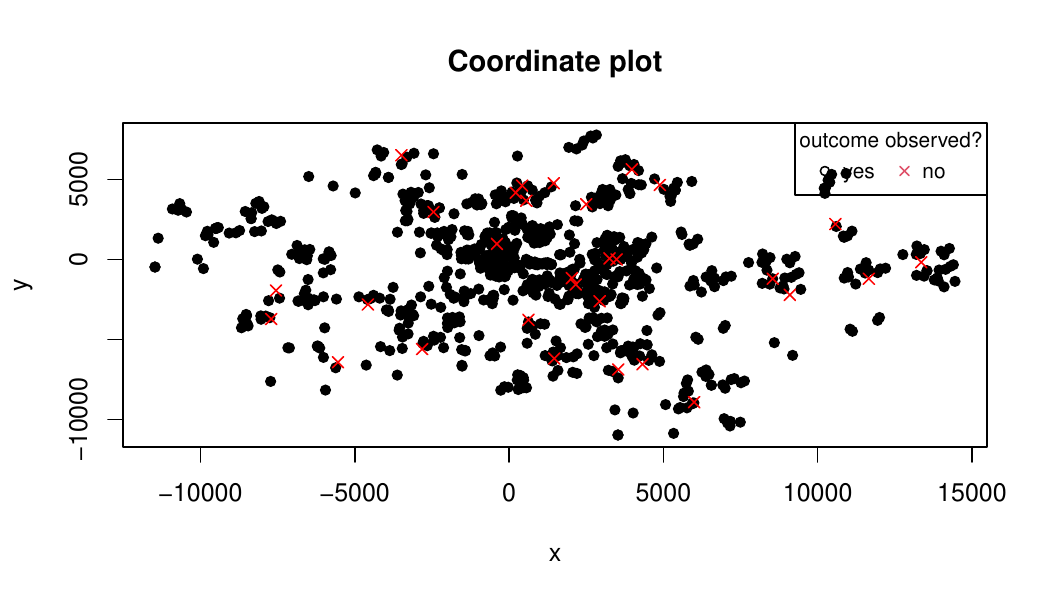} 
\caption{Coordinates plot for outcome \texttt{inc} with 30 random missing values} 
\label{inc}
\end{figure}

To illustrate the display of missing values we create a new matrix with the coordinates and the variable \texttt{inc} and then insert 30 random missing values.
In Figure \ref{inc} the missing values are marked with red crosses. As expected no spatial pattern is visible.

Further information on the distribution of pairwise Euclidean distances is provided by the function \texttt{distance.info()} which calculates
\begin{itemize}
\item the distance matrix containing all pairwise Euclidean distances,
\item the set of all pairwise Euclidean distances  where duplicate values due to symmetry are deleted.
\end{itemize}
Moreover, \texttt{distance.info()} displays the following descriptive statistics:

\begin{itemize}
\item a histogram of the Euclidean distances,
\item  minimum, 1st quartile, median, mean, 3rd quartile and maximum of the Euclidean distances.
\end{itemize}

\begin{figure}[t]
\begin{kframe}\begin{verbatim}
distance.info(birth)
\end{verbatim}
\end{kframe}
\centering
\includegraphics[width=\textwidth]{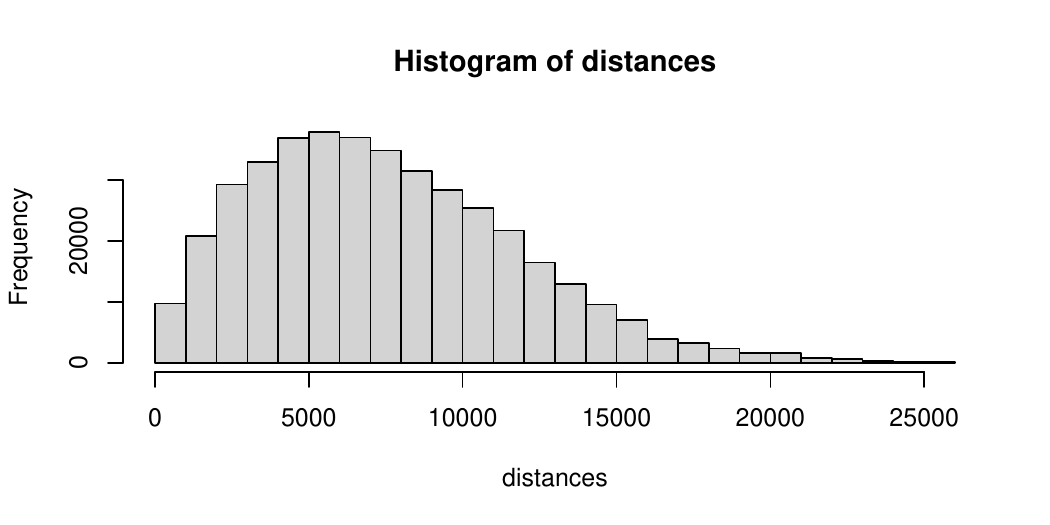} 
\begin{kframe}\begin{verbatim}
##  Summary of distance set: 
##    Min. 1st Qu.  Median    Mean 3rd Qu.    Max. 
##       0    4244    6970    7506   10221   25963
\end{verbatim}
\end{kframe}
\caption{Histogram of pairwise distances for the birth data.}
\label{hist}
\end{figure}

The output for the birth data is as follows and illustrated with the histogram shown in Figure \ref{hist}:
From the 815 409 pairwise distances total, 30 570 are of less than 2 000 meters and will be used for modelling of the local spatial correlation structure.

\subsection{Semi-variogram model fitting}\label{sec:modelling}
The function \texttt{vario.mod()}  enables the simultaneous output of multiple exponential semi-variogram models fitted for a range of maximal distances and bin numbers.
Thereby, the focus lies on the ability of the function to provide multiple estimation results depending on various specifications for the meta parameters \texttt{max.dist} and \texttt{nbins}.

It is advised to try out different values for both parameters and choose the model with the best fit. Commonly, the fit is evaluated by visual checks. An additional check can be performed by comparing the sample variance with the estimated variance according to the semi-variogram model $\hat{\sigma^2} = \hat{c_0} + \hat{\sigma_0^2}$ \citep{sauzet2021} by looking at the RSV.

The chosen maximal distance value specifies the subset of data pairs that are used for the semi-variogram estimation. Only data pairs with an Euclidean distance $\leq$ \texttt{max.dist} are taken into account.
For a first exploration, it might be useful to try a range of maximal distances to locate where the range might be situated.

\begin{figure}[t]
\centering
\begin{kframe}\begin{verbatim}
vario.mod(birth, max.dist = c(2000,1500,1000,500), nbins=13, shinyresults = TRUE)
\end{verbatim}
\end{kframe}
\includegraphics[width = 0.7\textwidth]{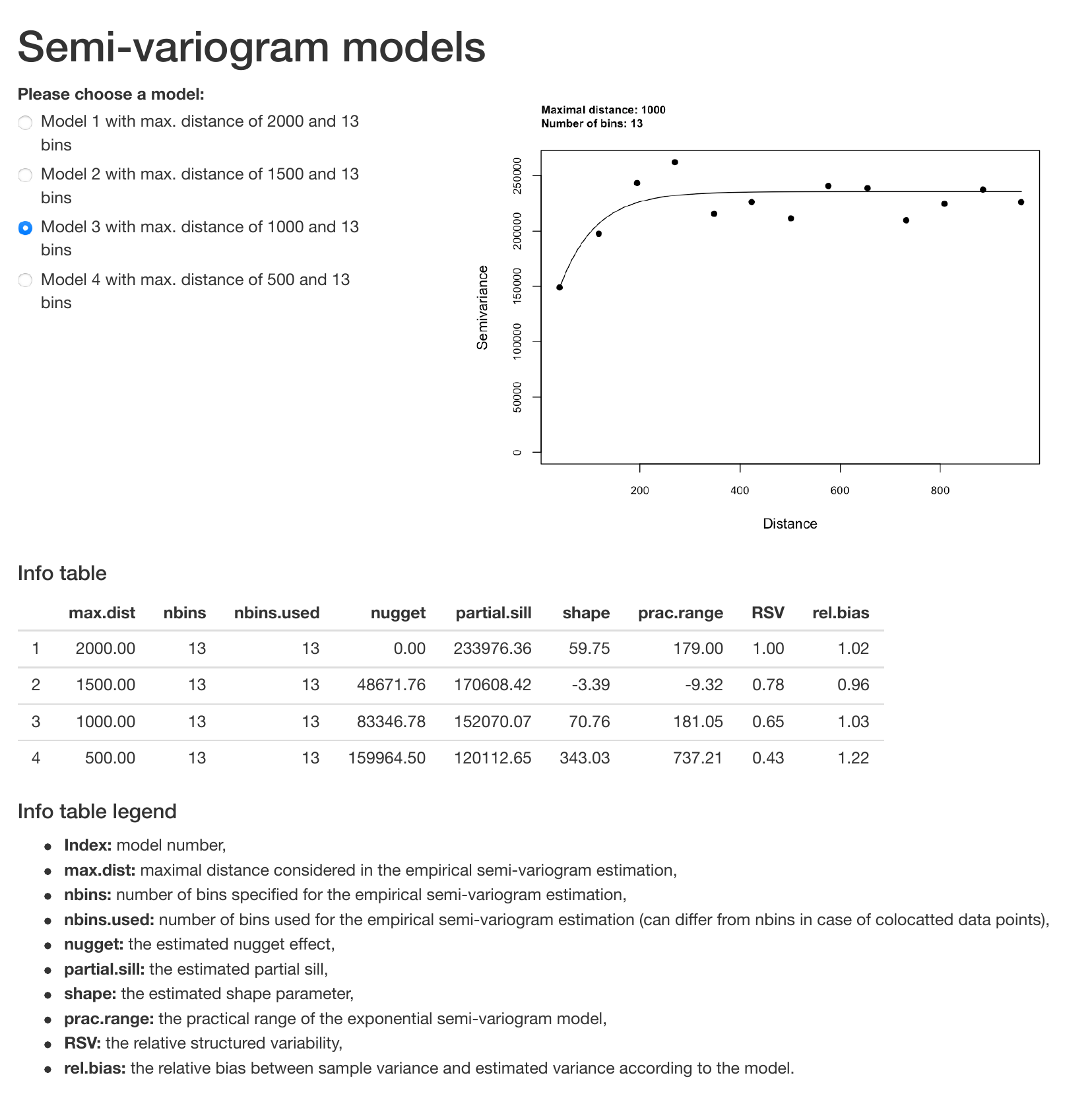}
\caption{Shiny output from \texttt{vario.mod()} for a range of \texttt{max.dist} choices.}
\label{birthweight}
\end{figure}

The code above Figure \ref{birthweight} will save a PDF file showing all fitted semi-variograms and will produce the {shiny} \citep{shiny} output shown in Figure \ref{birthweight}.

Each row of the printed output table (see Figure \ref{birthweight}) contains the estimated parameters of the exponential semi-variogram model with one of the stated maximal distances. More precisely, the table columns  contain:
\begin{itemize}
   \item \texttt{index}: model number,
    \item \texttt{max.dist}: maximal distance used in the estimation of the empirical variogram,
    \item \texttt{nbins}: number of bins specified for the empirical variogram estimation,
       \item \texttt{nbins.used}: number of bins used for the empirical semi-variogram estimation (can differ from nbins in case of colocatted data points),
   \item \texttt{nugget}: the estimated nugget effect $\hat{c_0}$,
  \item \texttt{partial.sill}: the estimated partial sill $\hat{\sigma}_0^2$,
   \item \texttt{shape}: the estimated shape parameter $\hat{\phi}$,
   \item \texttt{prac.range}: the practical range of the exponential model,
   \item \texttt{RSV}: the relative structured variability,
  \item \texttt{rel.bias}: the relative bias between the sum of the estimated partial sill and nugget and the sample variance (which theoretically are the same).
\end{itemize}

\begin{figure}[t]
\centering
\begin{kframe}\begin{verbatim}
vario.mod(birth, max.dist = c(1000, 800, 600), nbins=13, shinyresults = TRUE)
\end{verbatim}
\end{kframe}
\includegraphics[width=0.7\textwidth]{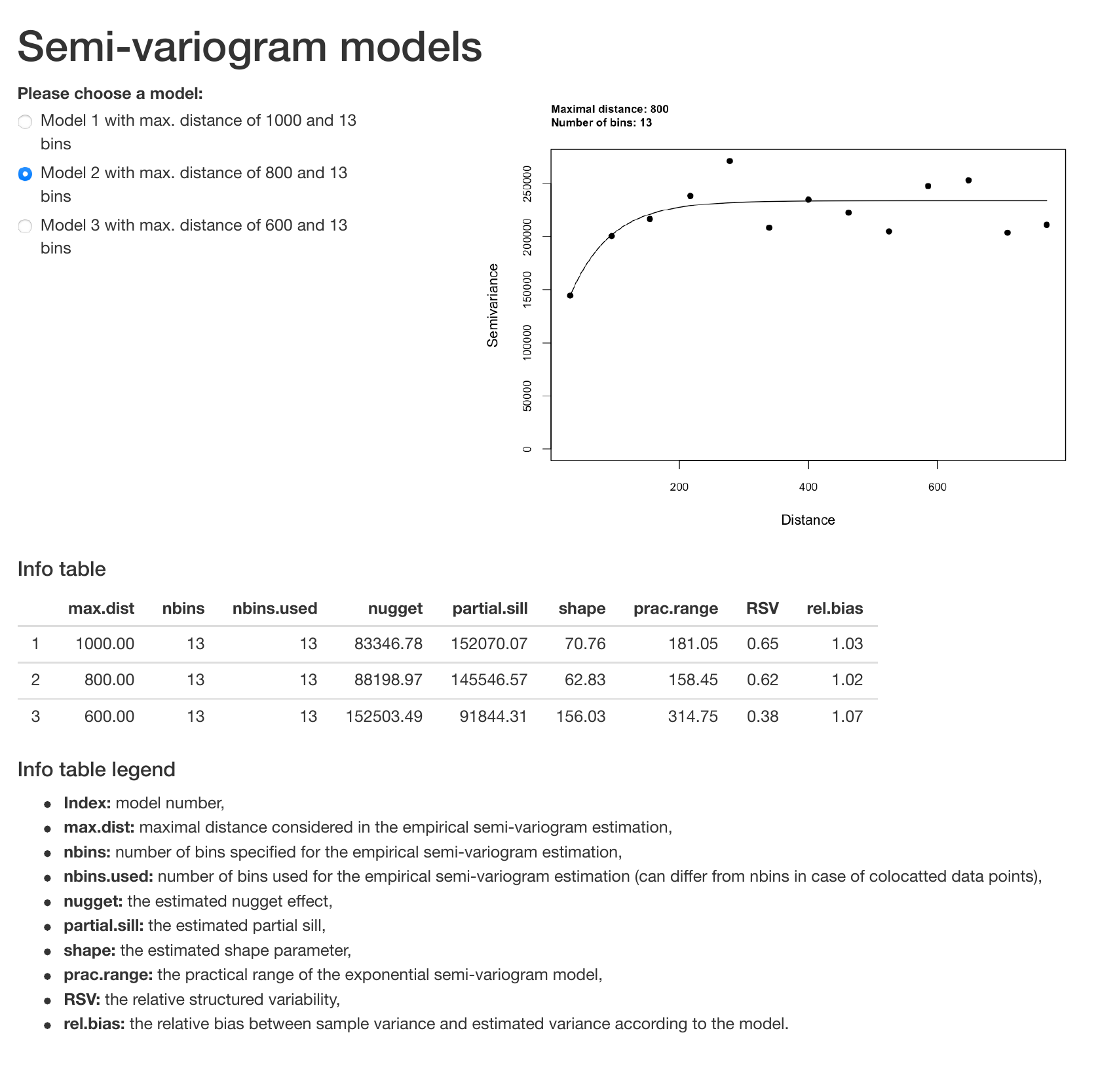}
\caption{Shiny ouput from \texttt{vario.mod()} for a finer range of \texttt{max.dist} choices.}
\label{birthweight2}
\end{figure}

A maximal distance of 1000 meters seems to provide the best fit among the options tried and we can now refine the analysis by considering a grid of smaller maximal distances
leading to the output presented in Figure \ref{birthweight2}.
Because a maximal distance of 800 meters provides the best fit for the exponential model with a low RB and a good visual fit, we further investigate the effect of the number of bins for a fixed maximal distance of 800.

The \texttt{nbins} parameter specifies the number of lags of the empirical semi-variogram to be estimated. On the one hand, a high number of lags might lead to a small within-lag sample size and thus to an unstable estimate. On the other hand, choosing a number of bins that is too small leads to a model that does not detect a spatial correlation structure at all. To decide on one or multiple values for \texttt{nbins}, taking a look at the histogram plot of the pairwise distances (see Figure \ref{hist}) obtained by \texttt{distance.info()} may help.

\begin{figure}[t]
\centering
\begin{kframe}\begin{verbatim}
vario.mod(birth, max.dist = 800, nbins = c(11,12,13), shinyresults = TRUE)
\end{verbatim}
\end{kframe}
\includegraphics[width=0.7\textwidth]{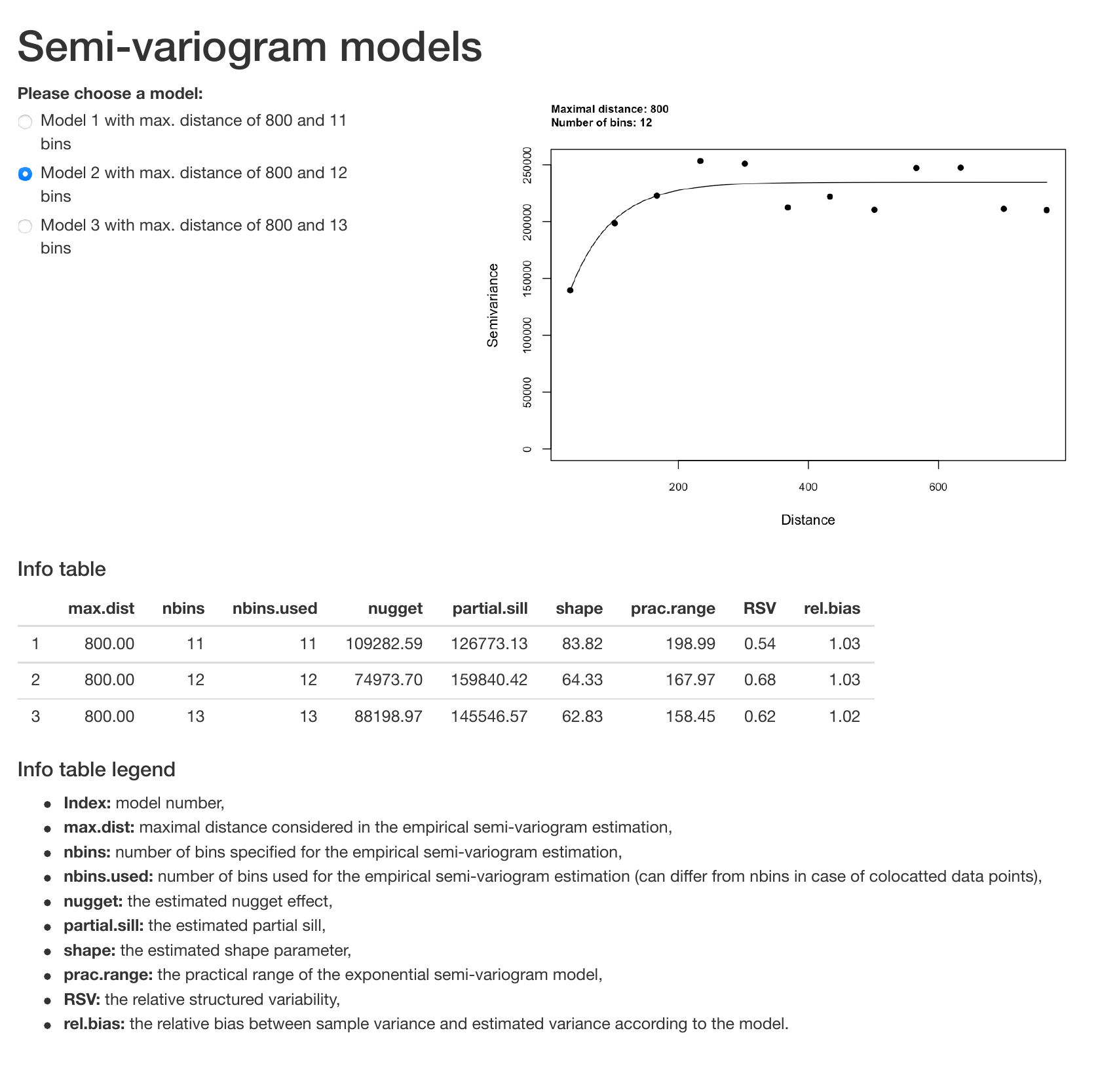}
\caption{Shiny output from \texttt{vario.mod()} for a range of \texttt{nbins} choices.}
\label{birthweight3}
\end{figure}

Trying out multiple \texttt{nbins} specifications as in the code above Figure \ref{birthweight3},
we obtain the output presented in Figure \ref{birthweight3}.
All models provide similar results but the third model with \texttt{max.dist = 800} and \texttt{nbins = 13} provides a slightly better fit and could be selected as the final model to capture the spatial effect on the health outcome \texttt{birthweight}.

\subsection{Modelling the spatial correlation structure of residuals}\label{sec:modellingres}

We want to investigate whether some or all of the observed spatial correlation structure can be explained by adjusting for predictors of birthweight.
Therefore, instead of modelling the correlation structure of a health outcome, the \texttt{vario.mod()} function can be used to model the spatial correlation structure of residuals from a (hierarchical) linear regression.
To do so, the studentized residuals from a (hierarchical) linear regression model are extracted via the
\texttt{vario.reg.prep()} function.

In the first step, we fit the following regression model and investigate the output:

\begin{knitrout}
\begin{kframe}
\begin{verbatim}
res <- lm(birthweight ~ datediff + primiparous + bmi, data = birth)
summary(res)
## 
## Call:
## lm(formula = birthweight ~ datediff + primiparous + bmi, data = birth)
## 
## Residuals:
##      Min       1Q   Median       3Q      Max 
## -1109.92  -274.10   -14.14   260.87  1373.98 
## 
## Coefficients:
##             Estimate Std. Error t value Pr(>|t|)    
## (Intercept) 3402.687     59.886  56.819  < 2e-16 ***
## datediff     -24.217      1.444 -16.773  < 2e-16 ***
## primiparous -108.669     28.424  -3.823 0.000141 ***
## bmi            6.551      2.506   2.614 0.009092 ** 
## ---
## Signif. texttts:  0 '***' 0.001 '**' 0.01 '*' 0.05 '.' 0.1 ' ' 1
## 
## Residual standard error: 415.9 on 899 degrees of freedom
## Multiple R-squared:  0.247,	Adjusted R-squared:  0.2445 
## F-statistic: 98.32 on 3 and 899 DF,  p-value: < 2.2e-16
\end{verbatim}
\end{kframe}
\end{knitrout}

All predictors are significant. 
Using the \texttt{vario.reg.prep()} function we now assign the studentized residuals of the regression model to the spatial coordinates and investigate which maximal distance provides the best exponential semi-variogram model fit for the studentized residuals saved in \texttt{v.prep}:
\begin{knitrout}
\begin{kframe}
\begin{verbatim}
v.prep <- vario.reg.prep(res, data = birth)
\end{verbatim}
\end{kframe}
\end{knitrout}

\begin{figure}[t]
\centering
\begin{knitrout}
\begin{kframe}
\begin{verbatim}
vario.mod(v.prep, max.dist = c(1000,800,600), nbins = 13, shinyresults = TRUE)
\end{verbatim}
\end{kframe}
\end{knitrout}
\includegraphics[width = 0.7\textwidth]{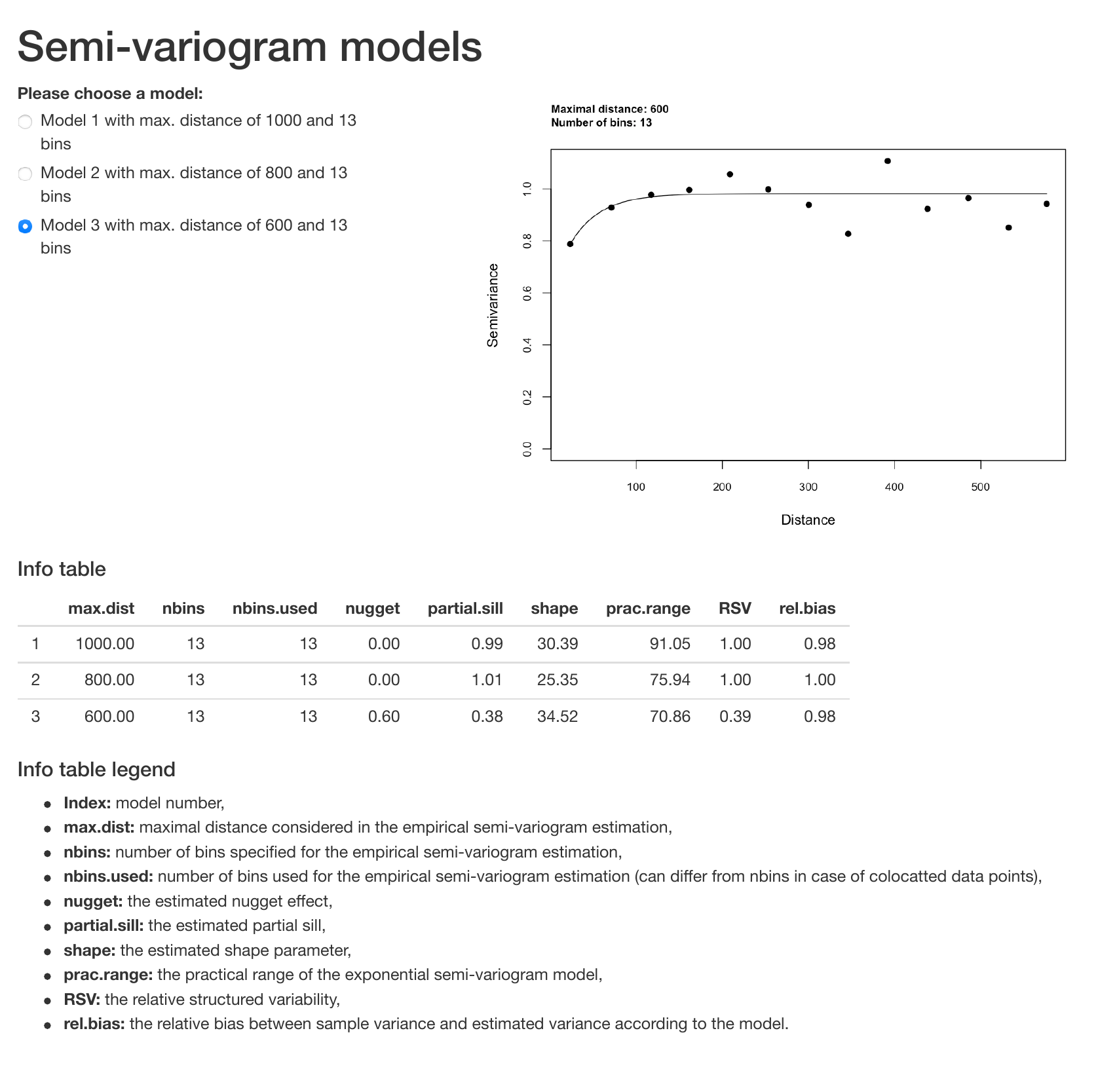}
\caption{Shiny output from \texttt{vario.mod()} based on residuals for a range of \texttt{max.dist} choices.}
\label{birthweight.adj}
\end{figure}

In terms of model fitting, we start at similar maximal distances to the ones chosen for the raw data (see Figure \ref{birthweight3}) and obtain the output depicted in Figure \ref{birthweight.adj}. The results point towards a reduced spatial correlation structure when controlling for various predictors with a well-fitting maximal distance of only 600 meters and much less regularity in the empirical semi-variogram.

Analogously to the unadjusted case (see Section \ref{sec:modelling}), we try out multiple \texttt{nbins} values fixing the maximal distance at 600 meters.

\begin{figure}[t]
\begin{knitrout}
\begin{kframe}
\begin{verbatim}
vario.mod(v.prep, max.dist = 600, nbins = c(11, 12, 13), shinyresults = TRUE)
\end{verbatim}
\end{kframe}
\end{knitrout}
	\centering
	\includegraphics[width = 0.7\textwidth]{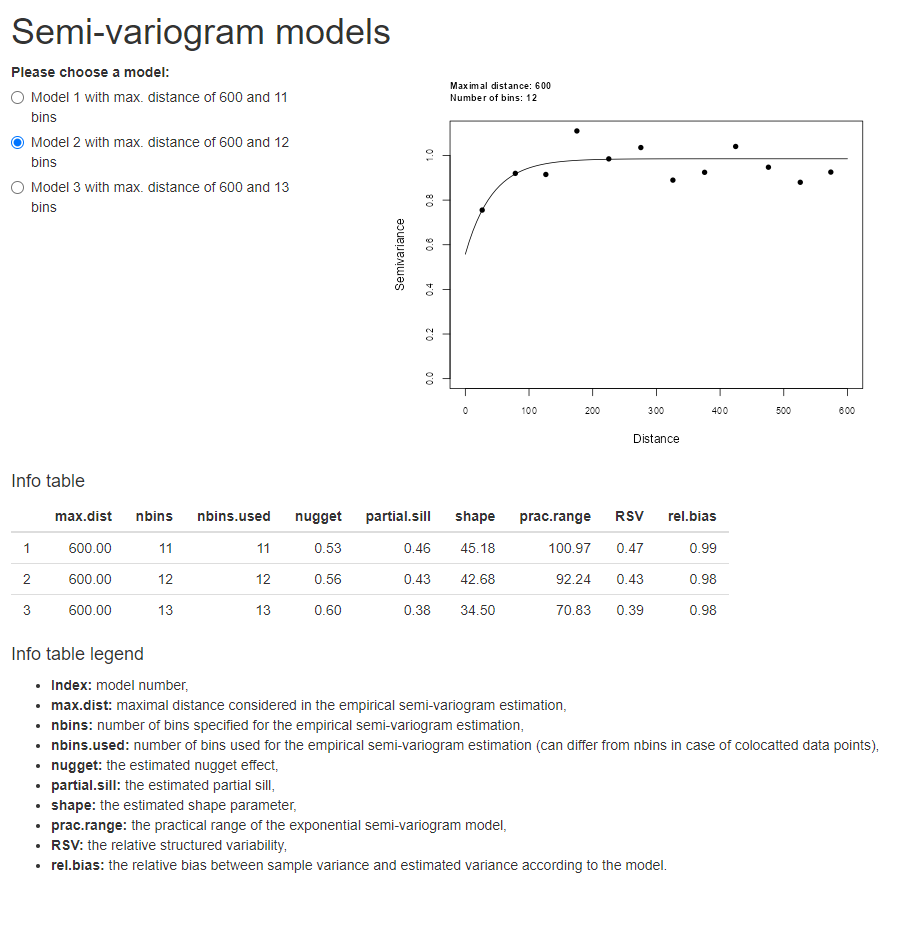}
	\caption{Shiny output from \texttt{vario.mod()} based on residuals for a range of \texttt{nbins} choices.}
	\label{fig:residualsnbins}
\end{figure}

Based on the resulting graphics and table (see Figure \ref{fig:residualsnbins}) we conclude that the models with \texttt{max.dist = 600} and \texttt{nbins = 12} (see Figure \ref{fig:residualsnbins}) or \texttt{max.dist = 600} and \texttt{nbins = 13} (see Figure \ref{birthweight3}) provide visually very similar results and fit the data best.

\subsection{Filtered bootstrap standard errors}
The function \texttt{par.uncertainty()} provides filtered bootstrap standard errors for all three exponential model parameters. Standard errors are important for conducting proper inference \citep{bard1974}. Moreover, they can be helpful to get an impression about the reliability of the estimated model and provide an objective tool to compare two or more models which seem to provide an equally good fit when evaluated visually as demonstrated in the last section \ref{sec:modellingres}.

Because the execution of the filtered bootstrap algorithm can be time consuming (depending on the sample size and number of bootstrap repetitions), the \texttt{par.uncertainty()} function is not called automatically within \texttt{vario.mod()} such that the bootstrap is not executed for all estimated models. The choice which model estimate should be completed by estimating standard errors, is left to the user by selecting the fitted model of interest in the \texttt{vario.mod()} outcome by its number in the \texttt{par.uncertainty()} option \texttt{mod.nr} and thereby saving execution time.

Due to the randomness in the bootstrap component within the standard error calculation, repeatedly estimating the standard errors for one fixed semi-variogram model will vary slightly.  
If the filtered bootstrap results are wished to be reproducible, a seed has to be set prior to the application of the \texttt{par.uncertainty()} command.

We save the \texttt{vario.mod.output} object containing the two best semi-variogram models of the residuals according to visual inspection:

\begin{knitrout}
\begin{kframe}
\begin{verbatim}
models <- vario.mod(v.prep, max.dist = 600, nbins = c(12, 13),
                    shinyresults = FALSE)
\end{verbatim}
\end{kframe}
\end{knitrout}
Based on that, we can estimate the parameter standard errors for model 1 (\texttt{max.dist = 600, nbins = 12}):
\begin{knitrout}
\begin{kframe}
\begin{verbatim}
unc1 <- par.uncertainty(models, mod.nr = 1, threshold.factor =3)
unc1$unc.table

##                 Estimate  Std. Error
## nugget effect  0.5575581   0.3177576
## partial sill   0.4275874   0.5015041
## shape         42.6818620 166.6720236
\end{verbatim}
\end{kframe}
\end{knitrout}

and model 2 (\texttt{max.dist = 600, nbins = 13}):
\begin{knitrout}
\begin{kframe}
\begin{verbatim}
unc2 <- par.uncertainty(models, mod.nr = 2, threshold.factor = 3)
unc2$unc.table
					
##                 Estimate  Std. Error
## nugget effect  0.5996697   0.3172093
## partial sill   0.3825593   0.5561611
## shape         34.5024985 203.0448548}
\end{verbatim}
\end{kframe}
\end{knitrout}

According to the above R output the standard error estimates of model 1 appear to be slightly smaller than the standard error results of model 2 confirming our impression that the models are very similar. The uncertainty quantification in terms of standard error values allows us to make an objective comparison and encourage us to select model 1 as the final model.


\section{Conclusion}

The R package {EgoCor} offers a range of functions to explore which empirical semi-variogram metaparameters (maximal distance and number of bins) result in the best-fitting exponential semi-variogram model. The graphical display and summary tables provided by the package facilitate the comparison of fitted semi-variogram models for multiple chosen metaparameters. Moreover, the package provides an implementation of the filtered bootstrap method proposed by \citet{dyck2023parameter} enabling standard error estimation for exponential semi-variogram model parameters that can be used for more formal model comparisons beyond the visual evaluation of the fitted semi-variogram.

Due to the user-friendly interface, practitioners can conveniently use the package to apply the concept of correlation neighbourhood suggested by \citet{sauzet2021} or use it to guide the choice of a suitable kriging model \citep{finne2024} and potentially more.


\section*{Computational details}

The results in this paper were obtained using
R~4.2.1 with the
{EgoCor}~1.2.0 package. R itself
and all packages used are available from the Comprehensive
R Archive Network (CRAN) at
\url{https://CRAN.R-project.org/}.



\newpage

\bibliography{refs}

\newpage

\begin{appendix}

\end{appendix}

\end{document}